\documentstyle[11pt,newpasp,twoside,epsf]{article}
\def\edcomment#1{\iffalse\marginpar{\raggedright\sl#1\/}\else\relax\fi}
\reversemarginpar

\begin{document}
\title{Disks and Planets in Binary Systems}
 \author{Wilhelm Kley \& Andreas Burkert}
\affil{Max-Planck-Institut f\"ur Astronomie,
K\"onigstuhl 17, D-69117 Heidelberg, Germany}

\begin{abstract}
The star formation process in molecular clouds usually leads
to the formation of multiple stellar systems, mostly binaries.
Remaining disks around those stars may be located around
individual stars (circumstellar disks) or around the entire binary system
(circumbinary disk).
We shall briefly review the present observational evidence for both types of
disks in binary stars, in particular the properties of circumbinary disks.

We then present recent results of the theoretical modeling of the collapse
and fragmentation of gravitationally unstable molecular cloud cores and their
implications for binary and disk formation, and discuss
the dynamical influence of the binary companions  on disk truncation and gap 
formation.
The presence of binaries may have profound influence on the process
of planet formation as well. We present results on the stability and
evolution of orbits of planets in disks around binaries.

\end{abstract}
\section{Observational Properties}
\label{obs_properties}
This contribution is dealing primarily with theoretical aspects, however
for completeness, we summarize first some basic observational features
of disks and planets in binary stars. An excellent recent review on this
topic has been given at PPIV by Mathieu et al. (2000).
\subsection{Binary Stars}
The majority of stars is found in binary or multiple systems.
To quantify the occurrence of multiplicity one usually defines the
{\it Binary Frequency} (BF) as
\[  BF = \frac{Number~~of~~multiples}{Total~~Number~~of~~Systems} \]

The seminal study of Duquennoy \& Mayor (1991) found for field stars 
(not members of a stellar cluster) having an age
larger than 1Gyr a binary frequency of $BF=60\%$.
The distributions of the orbital elements of binary stars
such as a mass ratio $q$, 
periods $P$, or eccentricities $e$ are typically very broad and they have 
mean values of
\[   <q> = 0.3 \]
\[     <\log(P[days])> = 5.0 \]
\[   <e> = 0.5 \]
For pre main-sequence stars one finds a bimodal binary frequency. Dense
clusters in giant molecular clouds, such as the Trapezium cluster in
Orion (with a stellar density of over 2000 stars per $pc^3$), tend to have 
a binary frequency similar to the field value, ($60\%$).
In low stellar density clusters (such as the Taurus or Ophiuchus T Tauri
associations) the binary frequency is
much higher than the field value, about $80-100\%$ 
(Ghez et al. 1993, Leinert et al. 1993).
This difference in the $BF$ may be the result of several mechanisms the most
important of which are (see Mathieu et al. 2000):
a) observational effects due to incompleteness may lead to a wrong 
determination of the BF, b) evolutionary effects may change the BF with time,
c) in higher density clusters more frequent stellar encounters may lead
to the disruption of systems and a lowering of the BF, and d) the BF
may just depend on the environment, such as the temperature or density
of the cloud core.

From these results one may infer that most of the field stars originated 
in dense clusters.
\subsection{Disks in Binaries}
The classical signatures for the presence of disks around stars 
are an excess in IR to mm radiation, Balmer and forbidden lines, polarization,
and optical veiling, see contribution by L. Hartmann (this volume). 

Disks in binary star systems may be divided into two main configurations
1) In the case of a {\it Circumbinary Disk} the complete
binary system is surrounded by a disk which may have in inner clearing.
2) If the two disks surround each component of the binary individually
the situation is similar to single stars and one refers to this 
configuration as {\it Circumstellar Disks}

The {\it mm-flux} from systems gives an indication for the mass of the disk.
For binaries with larger separations ($d> 100AU$)
and for spectroscopic binaries ($d<1 AU$) the mm-flux is similar to the field
value. This indicates that in the first case (large separations) the two
stars are surrounded individually by circumstellar disks behaving just as
disks around unperturbed single stars.
In the second case (very close stars) the system is surrounded by one
circumbinary disk which, when observed form far away, resembles 
a disk around one single star.

In binaries with intermediate orbital
separations $d$ between 1 and 100 $AU$ the mm-flux
is mostly undetected, which may indicate that in this case the
binary system has neither extended circumstellar disks,
as a consequence of the gravitational disturbance of the companion,
nor luminous circumbinary disks.
However, there are classical TTau signatures for disks with
$d=10-100AU$. 

Additionally, speckle and mid-IR observations give
evidence for circumstellar disks around the stars HK~Tau, HR~4796A, or
GG~Tau3. 

The most prominent observational example of a circumbinary disk is presented
by GG~Tau, where the disk has been directly imaged in the CO 1.3mm
line (Guilloteau 1999), in the J-Band at $1.2 \mu m$ (Roddier et al. 1996)
and in the optical. The central binary system consists of two
stars with a total mass of $1.3 M_\odot$ at a separation of $d=45AU$,
which are surrounded by a disk  
with an inner radius of $r_{disk}=180AU$.
The inferred mass of the disk (ring) is about $0.12 M_\odot$.
Additional systems with imaged circumbinary disks are UY~Aur and BD+31$^o$643.

In several spectroscopic binaries (eg. UZ Tau, DQ Tau or GW Ori) 
mm-observations have indicated disk masses of about $0.03$ to $0.05 M_\odot$
At the same time there is evidence for inner disk clearing 
(lack of near IR-emission), but there are signs for hot gas near the
stars which may indicate the flow of material through the cleared
inner disk (leaking gaps).
 
An interesting example of the simultaneous
occurrence of circumstellar and circumbinary disks
in a quadruple system is given by UZ Tau (Jensen et al. 1996). 
\subsection{Planets in Binaries}
Among the known 32 extrasolar planets around main sequence stars three
(16 Cyg B, 55 Cnc, $\tau$ Boo)
are definitely known to be in binary systems with relatively wide 
orbital separations (see table 1).
Additionally, in one system (Gl 86) there is some evidence for a stellar 
companion with a distance which may be as close as $10 AU$.
\begin{table}
\caption{The Properties of known planets in binary stars}
\begin{center}
\begin{tabular}{cccc|c}
   Parameter  &   16Cyg B & 55 Cnc & $\tau$ Boo & Gl~86\\
\hline
  Binary $d$ (AU)  &  700  &  1150  &  240 & $> 10 $\\
  M sin$i$ ($M_J$)  &  1.5  &  0.84  & 3.87 & 4 \\
  Planet a  (AU)   &  1.72 &  0.11  & 0.046 & 0.11\\
  Planet e         &  0.63 &  0.05  & 0.02 & 0.04 \\
\hline
\end{tabular}
\end{center}
\end{table}
It has been argued that the high eccentricity of 16 Cyg B may be caused
be the presence of a companion star.
However,
as the ratio of distance $d$ of the secondary over the semi-major axis $a$
of the planets falls into the range 
\[  \frac{d}{a} =  400 - 10000 \]
the companion has basically no influence on the eccentricity of the planet.
\section{Star Formation}
The observational evidence strongly suggests that
binary formation is closely related to the star formation
process. Stars form in dense molecular cloud cores (Myers, 1985, 1987)
with scales of order 0.1 pc, masses in the range of a few solar
masses, number densities of order 10$^4$ - 10$^5$ cm$^{-3}$ and 
temperatures of 10 K. In general, pre-stellar cores appear
to be elongated and irregular. They are centrally condensed with
Gaussian or power-law density distributions (Ward-Thompson et al. 1994;
Henriksen et al. 1997). Rotation has been detected in about 50\%
of all cases with values for the angular velocity ranging between
$\Omega \approx 10^{-13} - 10^{-15}$ s$^{-1}$ (Goodman et al. 1993).

The initial conditions for protostellar collapse could be strongly
affected by magnetic fields, which can support the cores against
collapse. In magnetically subcritical cores, ambipolar diffusion 
(Shu et al. 1993) leads to central contraction
and the formation of an $r^{-2}$-density distribution at which point
the core experiences an inside-out collapse. If the initial conditions
are produced in this manner, numerical simulations
have shown that the core would not fragment, leading to the formation
of single stars (Myhill \& Kaula 1992)
\begin{figure}[h]
\plotone{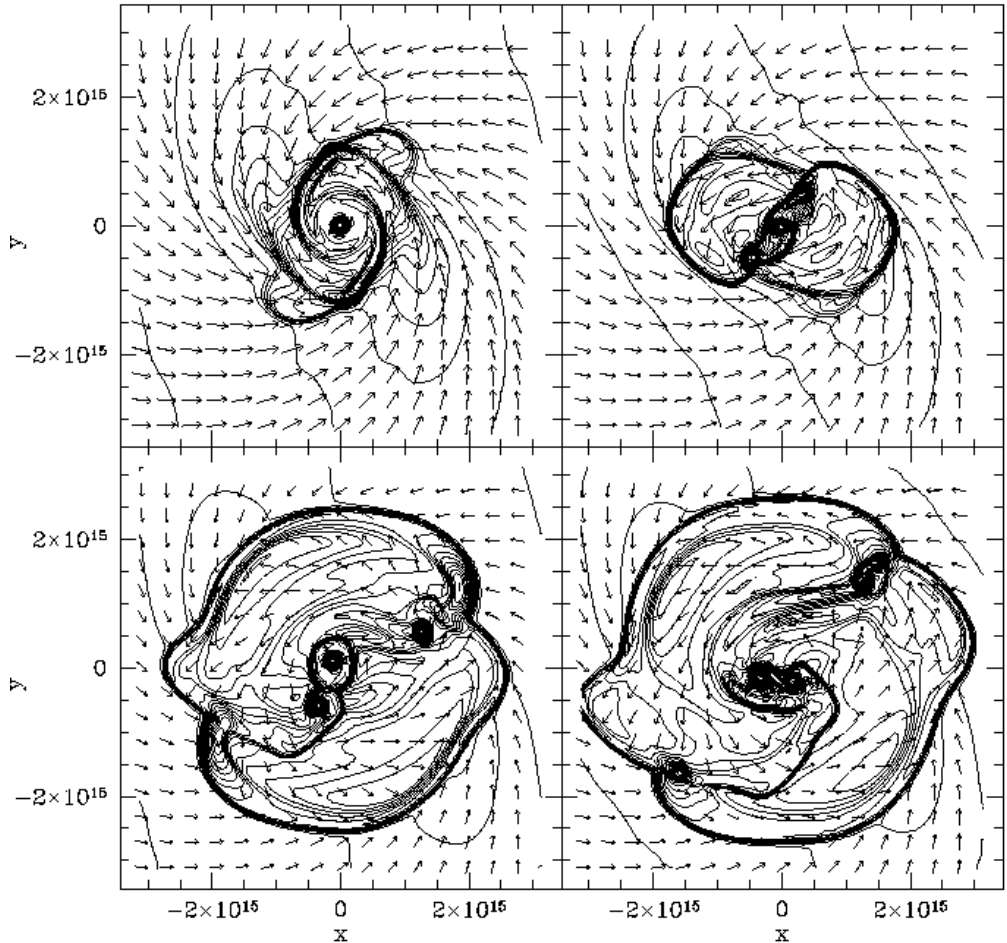}
\caption{Evolution of a collapsing gas cloud with an initial 
power-law density distribution. Logarithmically spaced contours 
of equal density and velocity
vectors in the equatorial plane are shown for the innermost region
(outer cloud radius: 5 $\times 10^{16}$ cm). The maximum density 
is $10^{-10}$ g cm$^{-3}$, the maximum velocity is
$3 \times 10^5$ cm s$^{-1}$.}
\label{fig.starformation}
\end{figure}
Binary formation results from the fragmentation of magnetically supercritical
cores.  A wide variety of (often idealized) initial conditions has been
explored numerically, adopting 3-dimensional hydrodynamical codes,
for example, a sphere (constant density or centrally condensed)
in uniform rotation with a density perturbation
in the form of an m=2 mode (Burkert \& Bodenheimer 1993, 1996;
Truelove et al. 1997), or elongated cylindrical, rotating clouds (Boss 1996, 
Bonnell et al. 1992). Most of these calculations treated the isothermal
collapse phase. Radiation has been included in some simulations
adopting simplified approximations (Boss 1993, Myhill \& Kaula 1992).

In general, the resulting fragmentation process occurs in various modes.
The most simple cases are non-linear initial density perturbations which
lead to fragmentation into wide binaries with separations of
order 100 - 1000 AU during the early collapse phase (Bonnell et al. 1992,
Monaghan 1994). In this case, the result depends strongly on the adopted
initial conditions. 

Fig.~\ref{fig.starformation} shows the collapse
of a centrally condensed gas cloud
with a linear m=2 density perturbation (Burkert et al. 1997). 
In this case, the perturbation
has not enough time to grow and become non-linear during the early
gravitational collapse phase. The gas instead settles into a
rotationally supported, partly self-gravitating protostellar disk.
The subsequent evolution of the disk depends critically on its temperature
(Bate \& Burkert 1997, Boss et al. 2000).
If the transition from the optically thin to the optically thick regime
occurs before disk formation, fragmentation will be inhibited. Otherwise
strong spiral arms form  around a central condensation
which continually wrap-up, interact and reform
because of differential rotation. Ultimately, density maxima appear
at the ends of the arms, leading to a binary around the central object.
As higher angular momentum material falls in, spiral arms extend beyond
the triple system which rapidly breaks up into a hierarchical triple.
Further fragmentation occurs in the outer parts of the disk. 
The multiple system will ultimately break-up into single stars
and eccentric binaries with wide period distributions.
\section{Influence of Companions on Disks}
Let us first consider the case of a disk surrounding one star 
(circumprimary disk) which is orbited by another secondary star. 
The presence of a companion excites tidal waves in the disk which carry
angular momentum and energy
 (see eg. Lin \& Papaloizou 1995, 
and contribution by C. Terquem, this volume).
If the disk lies inside the orbit of the companion the wave will 
carry negative angular momentum because the companion has a smaller orbital
velocity than the disk material. Dissipation of the wave,
for example through shock waves, will slow down the matter in the disk
and lead to an inward drift of the material.
This truncates the outer edge of the disk.

In case of a binary star within a circumbinary disk, the torques created
by the binary lead to angular momentum transfer from the binary's orbit 
to the disk, and eventually to an inward truncation or clearing of 
the inner disk (gap formation).

The location of this (inner and outer)
truncation radius is determined by the balance
of viscous (gap-closing) and gravitational (gap-opening) torques.
In linear theory the gravitational potential is expanded into a
Fourier series and for each component the effect on the disk is determined.
For the general case of a companion on an eccentric orbit this 
analysis has been performed for circumprimary and
circumbinary disks by Artymowicz \& Lubow (1994).

\begin{figure}[t]
\plottwo{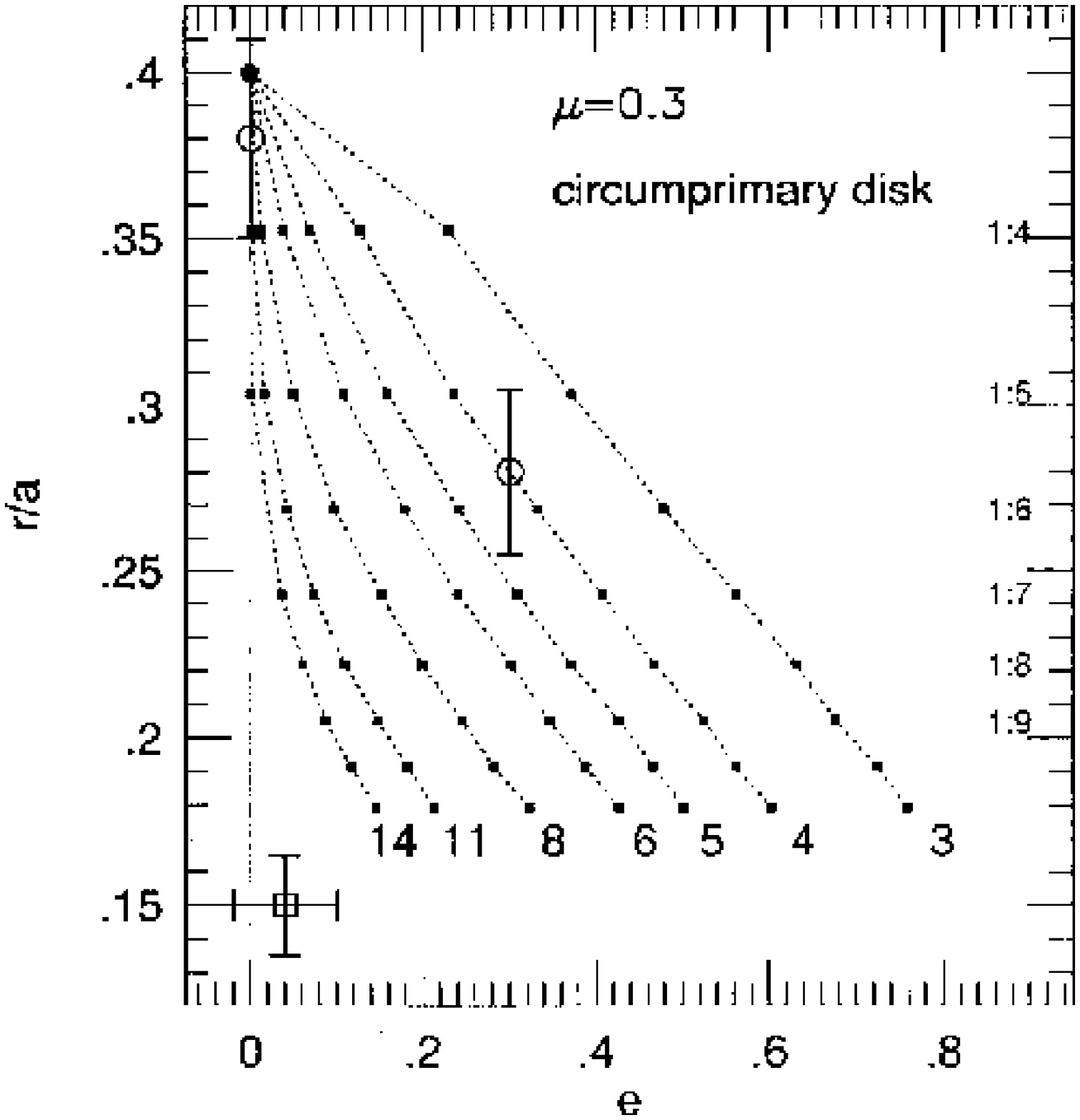}{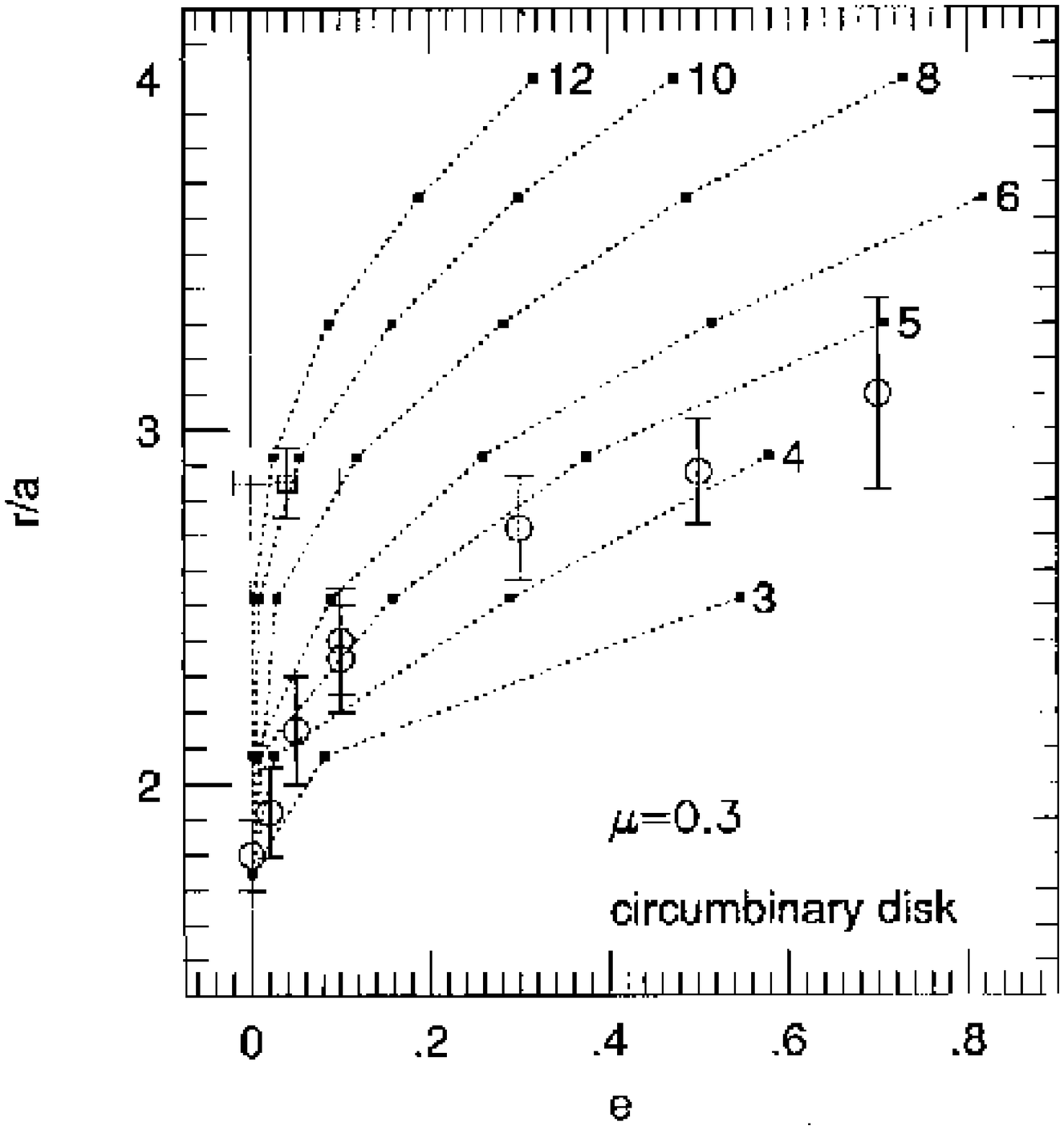}
\caption{
\label{fig.truncation}
The truncation radius of the disk in units of the semi-major axis
of the binary for a mass ratio $q=M_2/M_1$, and an eccentricity $e=0.5$.
Different values of the viscosity are labeled by the logarithm of
the Reynolds number in the disk.
Adapted from Artymowicz \& Lubow (1994).
}
\end{figure}

Their main results are displayed in Fig.~\ref{fig.truncation}.
The label $\mu = 0.3$ refers to the reduced mass $\mu = M_2/(M_1+M_2)$.
In case of the circumprimary disk a larger viscosity (lower
Reynolds number) leads to a larger truncation radius $r_t$ while for a
circumbinary disk it reduces the truncation radius.
For typical values of protostellar disks $Re\approx 10^5$, and for
typical binary parameter (see above, $q=0.5$, $e=0.5$) we obtain  
for the circumprimary disk $r_t/a=0.17$ and for the circumbinary disk
$r_t/a=3.0$.
\section{On the Formation of Planets in Binaries}
In this final section we consider the restrictions the presence of
a secondary companion has on the efficiency of planet formation. The secondary
acts as a perturber for the disk and alters directly the environment
in which planets form.
\subsection{Planetesimal disk}
The influence a secondary has on the growth process from planetesimals
to planetary embryos was studied by Heppenheimer (1978) and Whitmire (1999).
They considered a planetesimal disk at runaway phase where
collisions and subsequent merging
between different particles lead to a rapid growth.
The collisions are only non-destructive if the relative velocities 
$U_{rel}$ are smaller than a critical value $U_{crit}=100m/s$.

\begin{figure}[t]
\plotone{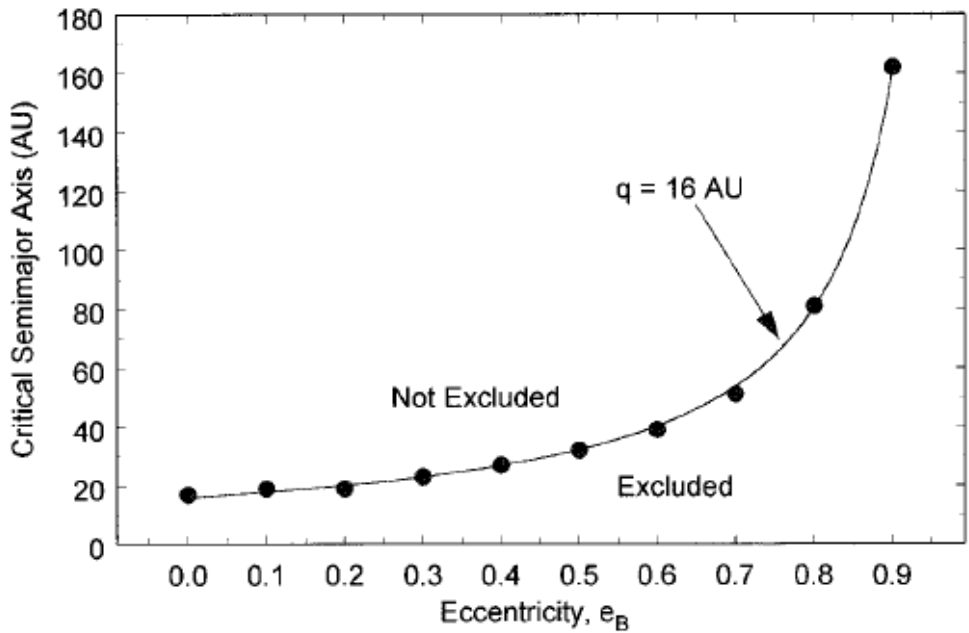}
\caption{
Critical semi-major axis for which collisions between planetesimals
are too large to allow for particle growth. The mass of the central
star was $1M_\odot$, the mass-ratio $q=1.0$, 
and the initial semi-major axis of the planetesimals was $1AU$.
Adapted from Whitmire et al. (1999).
}
\label{fig.whitmire}
\end{figure}

They integrated a 4-body system consisting of 2 stars and 2 planetesimals 
using a symplectic mapping procedure and an integration time
over a few $10^4yrs$.
For different physical parameter, varying the semi-major axis
of the binary stars, their eccentricities, mass ratios, and
the initial separation
in semi-major axis $\Delta a_0=0.001-0.01AU$ 
of the planetesimals, they analyzed  
if $U_{rel}$ was exceeding $U_{crit}$. They deduced a critical semi-major
axis below which the disturbing companion does not allow for 
sticking (growing) collisions.
Their main results are summarized in Fig.~\ref{fig.whitmire} where
the critical semi-major axis is plotted for an equal mass binary versus
eccentricity. If the separation of the binary falls below the curve 
the induced relative velocities are too large and planetesimals
cannot grow. The results are normalized to the initial mean separation
$\bar{a}=1 AU$ of the planetesimals.
The results of the numerical computations are given by the black dots,
and the solid curve gives the location where, for the given
eccentricity, the minimum separation (peri-astron) of the stars is exactly
16 AU.

The result implies that for a one-solar mass star the minimum distance
to an orbiting companion, such that planetesimals at 1AU can 
grow to larger bodies, must always be larger than 16 AU.
The critical semi-major axis scales only weakly,
$a_{crit} \propto (\bar{a}/AU)^{0.8}$, with
distance from the star.
\subsection{The case of L 1551}
Another path of studying the formation of planets in binary stars was taken
by A. Nelson (2000), who studied the interaction of two binary stars each of
which is surrounded by its own circumstellar disk. 
The physical motivation is based on the radio (VLA) observations
of the system L1551 (Rodriguez et al. 1998), which consists of a binary
system of two half solar mass stars separated by $50AU$. 

This system is modelled numerically by solving the full hydrodynamic equations
for the two disks using the method of Smoothed Particle Hydrodynamics (SPH),
where the continuum equations are modelled by an ensemble of interaction
particles. 
\begin{figure}
\plotone{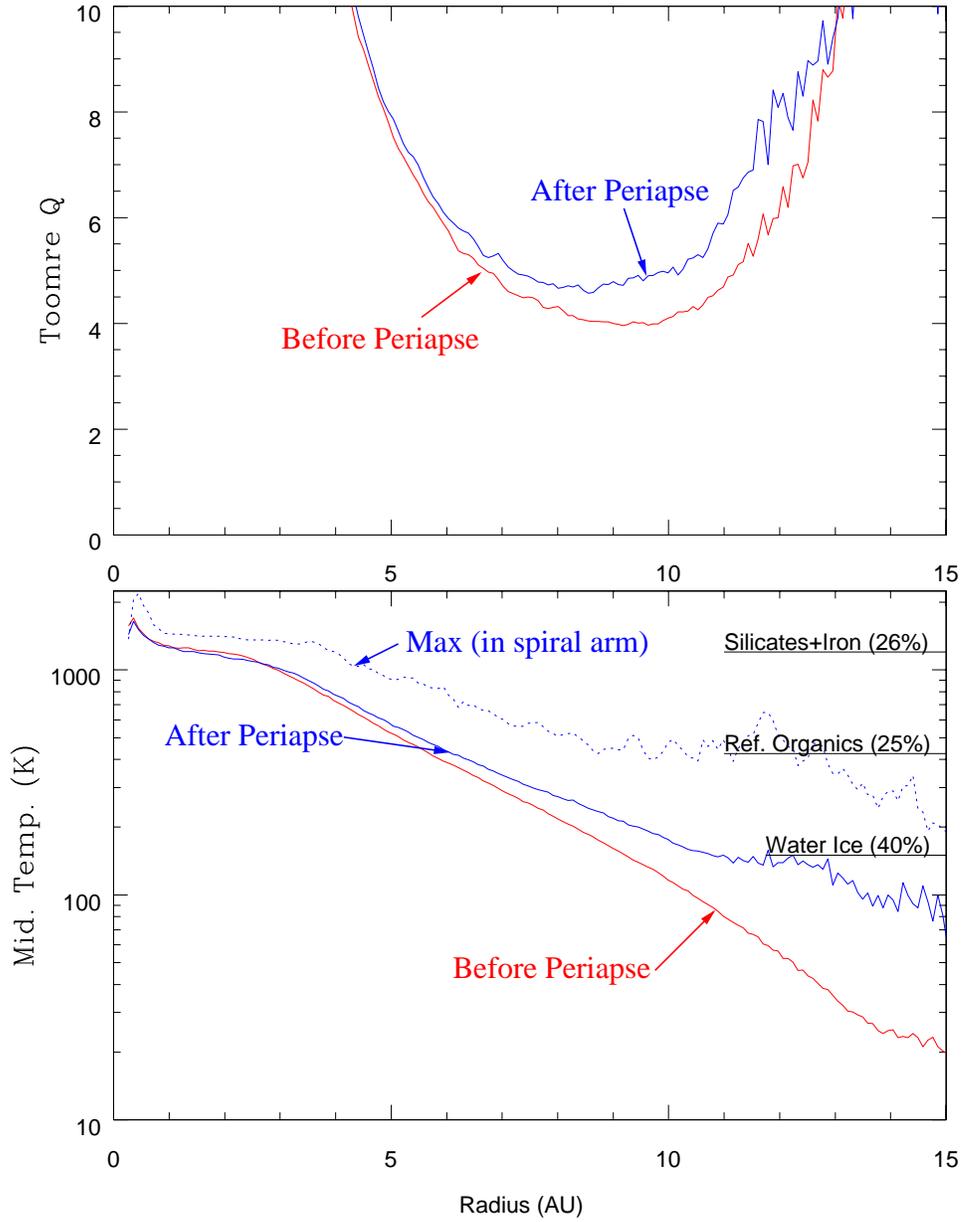}
\caption{
Azimuthally averaged Toomre-parameter $Q$ (top) and the midplane temperature
in the disk versus radius.
Indicated are the vaporization temperatures of various species.
}
\label{fig.andy}
\end{figure}

Each disk is modelled by 60,000 particles.
The model includes the self-gravity of the disks, an ideal equation of state,
dissipative heating and radiative cooling.

During the evolution the initially axisymmetric disks
(with respect to their central
stars) are strongly perturbed at the time of periapse. Spiral
waves are induced in them and they are tidally truncated.
At times of apoapse  the spiral features tend to disappear and the
disks become more axisymmetric again. 
These periodic changes of the distortions alter the internal structure
of the disk as well. The question arises what influence do these changes have
on the planet formation process.
Planet formation is believed to proceed along two different lines:
{\bf a})
Through gravitational collapse of a spiral structure.
Instability is given when the so called Toomre parameter
$Q$, which measures the importance of pressure versus gravity, becomes (locally)
smaller than unity. Then pressure forces are not sufficient to prevent
gravitational collapse.
{\bf b})
Through coagulation of solid material and subsequent gas accretion.
As can be inferred from Fig.~\ref{fig.andy} the $Q$-parameter is always
much larger than 1, indicating a gravitationally stable disk.
On the other hand, the temperatures in the disk during periapse
are so high that even
the most abundant species (water ice) cannot condense to form the seeds
for further growth of the planetesimals. 

Hence, from the computations one may conclude that, at least for these
chosen parameters of the disks, planet formation is inhibited in
binary stars. One has to keep in mind however, that the heating and cooling
of the disk is treated still approximative and the calculations are only
two-dimensional. For larger orbital separations of a few hundred $AU$ say,
the effects of the secondary become smaller and planet formation will not
be affected. 
\subsection{Stability of Orbits}
An important pre-requisite for the formation of planetesimals
is the long term stability of their orbits. Dvorak (1986) and 
Holman \& Wiegert (1999) have studied the restricted 3-body problem, in
which the planet is treated as a test particle in the potential of two
orbiting stars. As for this problem no analytic description 
of the orbital evolution of the planet exists they perform numerical
integrations using a symplectic mapping or Bulirsch-Stoer integrator.
The total integration time covers more than $10^4$ orbits of the planet.
The mass ratio of the stars, their eccentricity and semi-major axis are
varied. The critical distance $a_{crit}$ a planet must have to be on
a stable orbit is determined as a function of these orbital parameter
of the binary star.

\begin{figure}
\plottwo{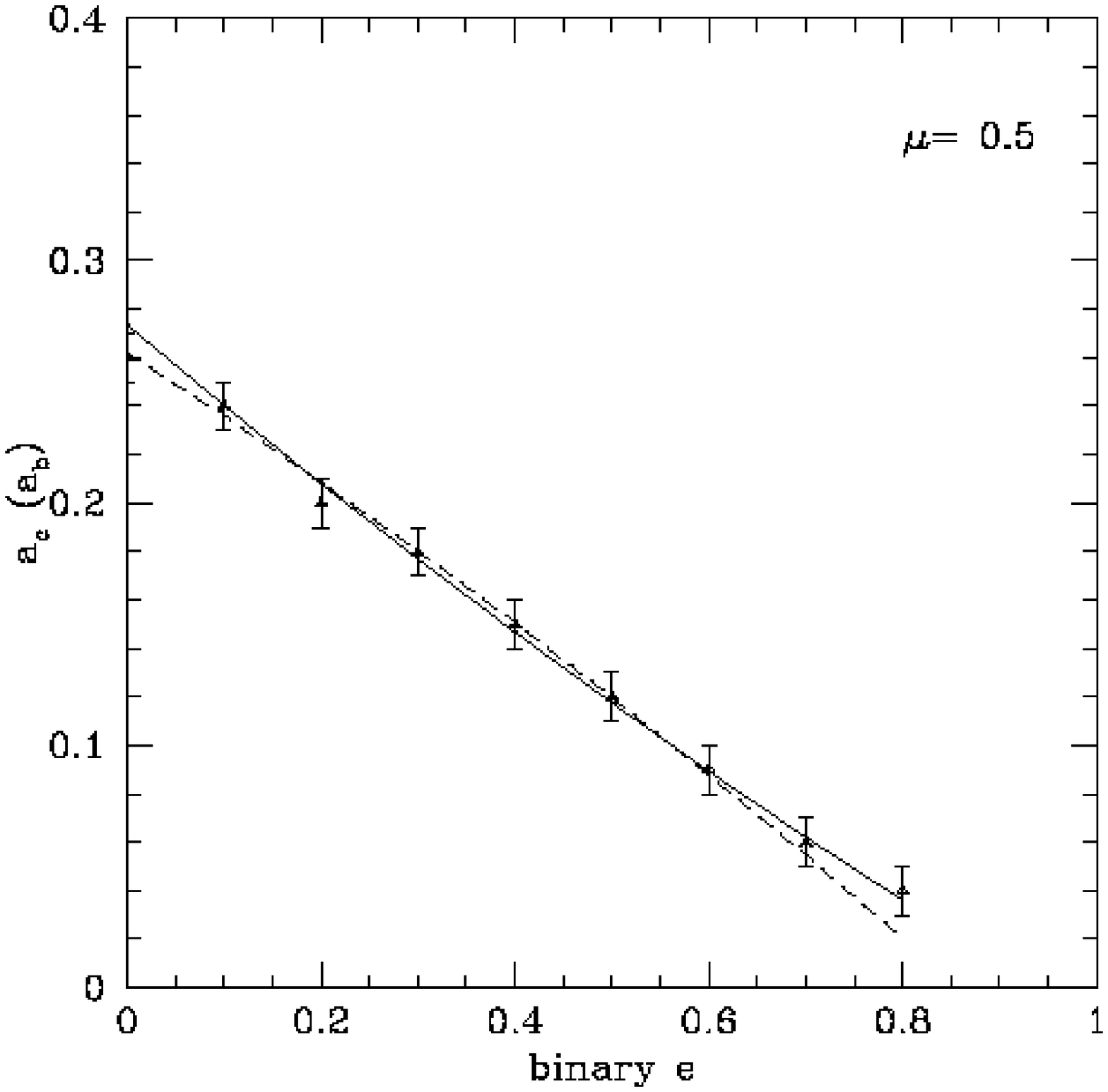}{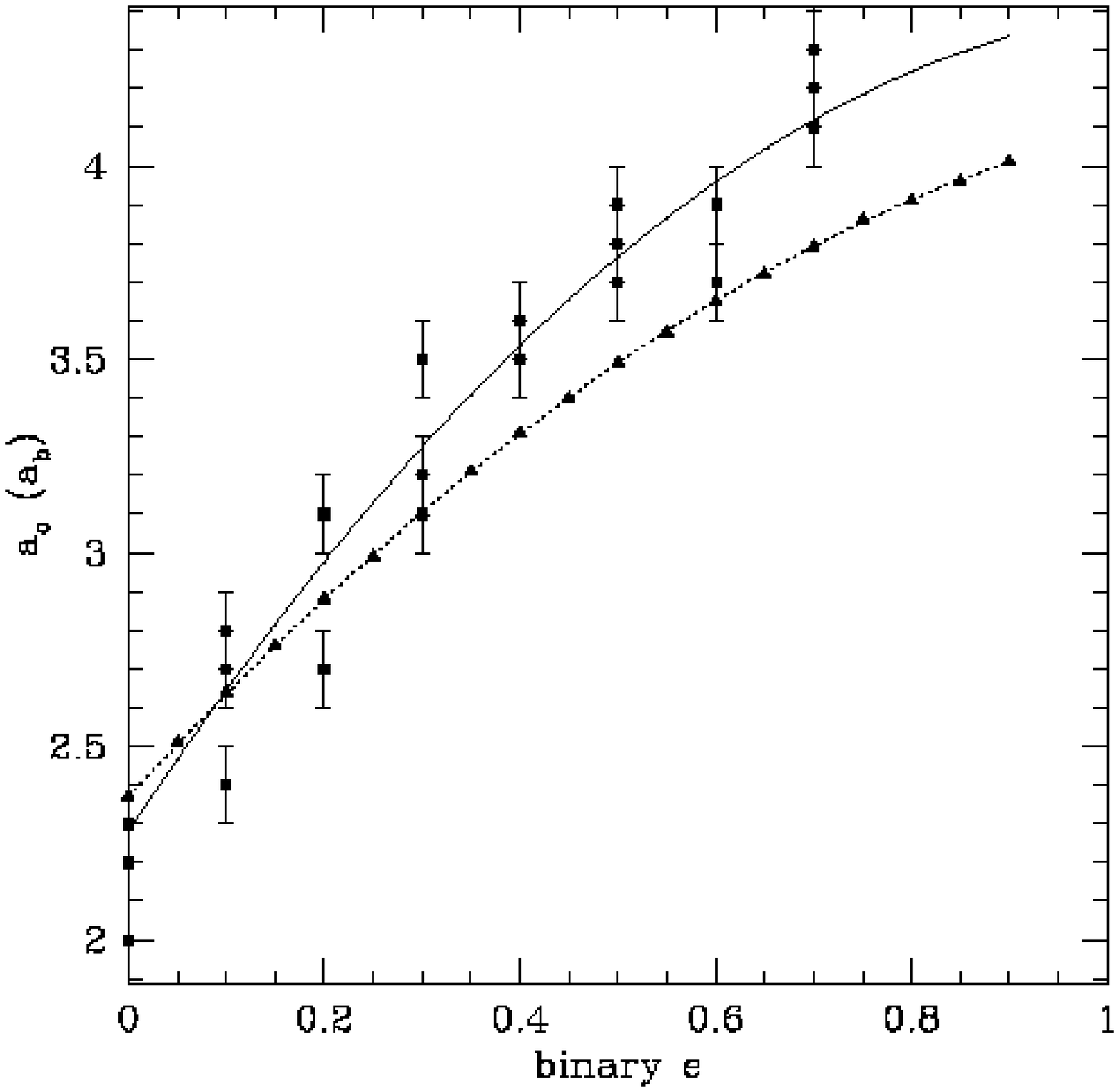}
\caption{
Critical semi-major axis for an equal mass binary. After Holman \& Wiegert
(1999).
{\bf Left:}
For S-type orbits. Separations smaller than the solid curve (which is
a fit the numerical data points) are stable.
{\bf Right:}
Evolution of the mass (bottom) and semi-major axis (top)
r P-type orbits. Separations above the solid curve are stable.
}
\label{fig.holman}
\end{figure}

Two different configurations of orbits have been studied. 
Following the designation of Dvorak (1986), 
the first are planetary or P-type orbits which are well outside the binary.
The second type studied are S-type orbits, where the planet orbits near one of
the stars, with the second star to be considered as a perturber.
For P-type orbits there exist a minimum $a_{crit}$ a planet must have to
be on a stable orbit around the binary. For S-type orbits there is a maximum
$a_{crit}$ such that the perturbations of the perturber remain finite.
The S-type orbits are certainly more relevant to the properties of the
observed extrasolar planets (see table~1).

The main results are displayed in Fig.~\ref{fig.holman} where,
for an equal mass binary, the 
critical distance (in units of the semi-major axis of the binary star 
system) is given as function of eccentricity.
For an average eccentricity of $e=0.5$ one finds for S-type orbits 
$a_{crit} \approx 0.14$. All observed extrasolar planets have semi-major
axis well below this limit. For P-type orbits the minimum distance a planet
must have from the binary is $a_{crit} \approx 3.6$. This value lies
beyond the inner truncation radius of a circumbinary disk ($r_{gap}=3.0$).

\subsection{The evolution of an embedded planet in a binary star}
Another approach to analyze the influence of a companion
on the formation of a massive planet has been taken by Kley (2000) who
studied the evolution of a massive planet still embedded in a 
protoplanetary disk.

A planet with the mass $1 M_{Jup}$ is placed initially on a
circular orbit around a $1 M_\odot$ star at a distance of $a_J =5.2 AU$. The
surrounding protostellar disk has a mass of $M_d = 0.01 M_\odot$ within
$1-20AU$, a surface density profile $\Sigma(r) \propto r^{-1/2}$, and a
Reynolds number of $10^5$.
The {\it secondary} has a given fixed mass of $M_2 = 0.5 M_\odot$ and
an eccentricity of $e_2 = 0.5$.
The semi-major axis $a_2$ of the secondary is varied from about 50 to 100
$AU$ for different models. To minimize initial disturbances, the secondary
is placed at apastron in the beginning.

During the evolution, the planet may accrete material from the disk and
increase its mass. At each timestep the gas density in the inner half
of the Roche lobe is reduced by a given fraction.
Initially, the disk is axisymmetric with respect to the primary star and has
a gap imposed to speed up the computations
(Kley 1999; Lubow, Seibert \& Artymowicz 1999).

\begin{figure}[t]
\plottwo{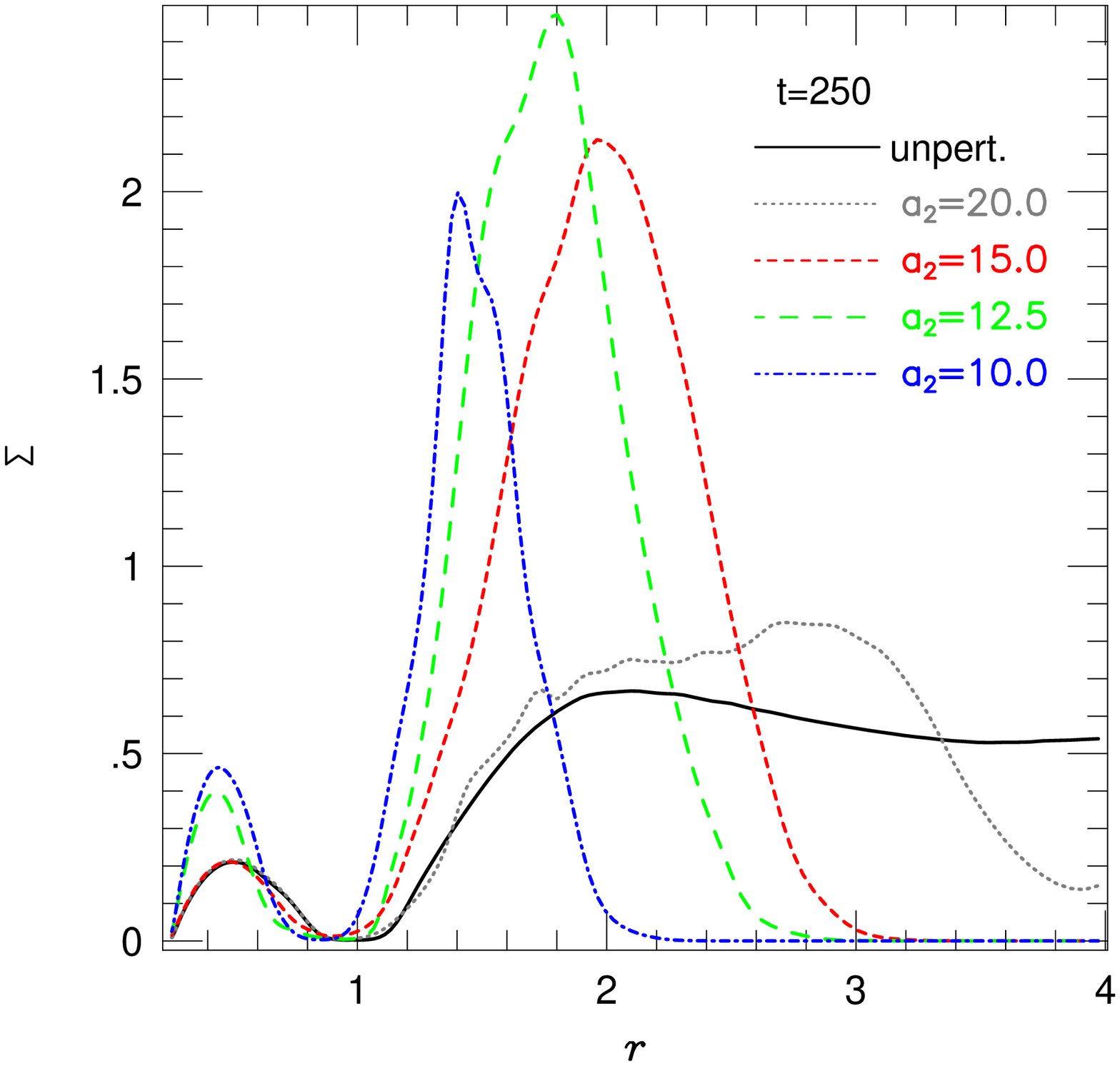}{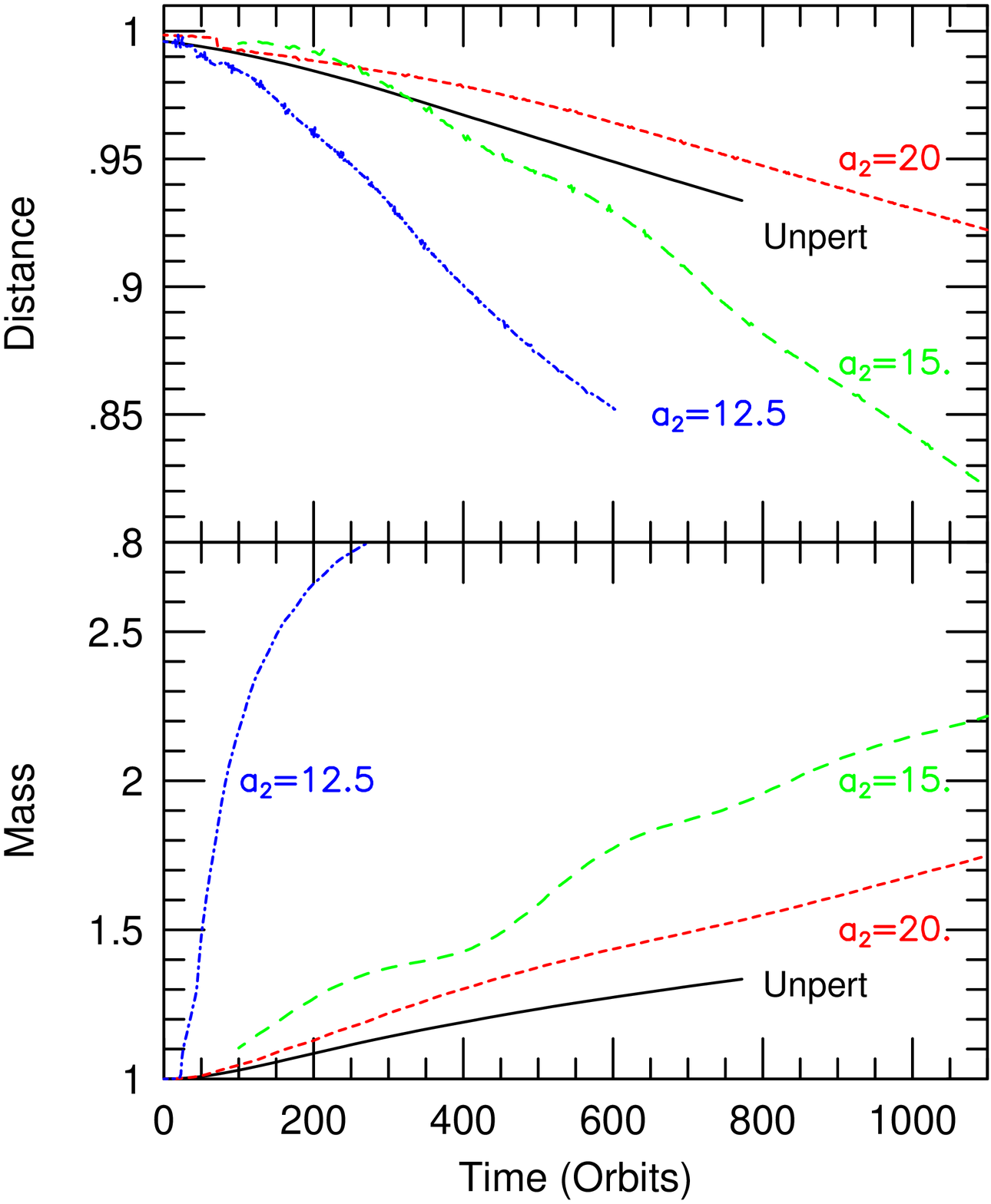}
\caption{
{\bf Left:}
Azimuthally averaged surface density for models with varying
distance ($a_2$) of the secondary, given in units of the
planetary distance ($5.2AU$), after 250 orbits of the planet.
The solid black line indicates the initial surface density profile.
{\bf Right:}
Evolution of the mass (bottom) and semi-major axis (top)
of the planet. 
The evolution of an unperturbed planet (i.e. no secondary star)
which is described in detail in Nelson et al. (2000) is given by the
solid lines (denoted "Unpert").
}
\label{fig.disktruncation}
\end{figure}

In the presence of the secondary the disk becomes truncated at the outer 
radius. At the same time the planet will truncate the disk from the
inner side such that the disk material will be confined to a narrow ring
(see Fig.~\ref{fig.disktruncation}). The obtained values for the outer
truncation radius agree favourably with the value $r_r/a=0.17$ as obtained
above.
The very steep density and pressure gradient near the planet lead to an
increased mass accretion rate onto the planet, and to a faster inward
migration, cf. Fig.~\ref{fig.disktruncation}.

The reduced migration time may then also reduce the growth of a
protoplanet and planet formation will again be inhibited by the presence
of a companion.
\end{document}